\documentclass[aps,prl,superscriptaddress,twocolumn,showpacs]{revtex4-1}
\usepackage{graphicx,amsmath,color,placeins}
\usepackage{amssymb,dcolumn}
\usepackage[colorlinks=true, linkcolor=blue, urlcolor=blue, citecolor=blue]{hyperref}
\let\hide\iffalse

\usepackage{newtxtext,newtxmath}

\usepackage[caption=false]{subfig}

\begin{document}
\title{Emerging kinetic-exchange for the enhanced metallic ferromagnetism in CrGeTe$_3$ under pressure}
\author{Jiaming Liu}
\author{Xuefeng Zhang}
\author{Hongbin Qu}
\affiliation{School of Physical Science and Technology, ShanghaiTech University, Shanghai 201210, China}
\author{Xiaoqun Wang}
\author{Hai-Qing Lin}
\affiliation{School of Physics, Zhejiang University, Hangzhou 310027, Zhejiang, China}
\author{Gang Li}
\email{ligang@shanghaitech.edu.cn}
\affiliation{School of Physical Science and Technology, ShanghaiTech University, Shanghai 201210, China}
\affiliation{\mbox{ShanghaiTech Laboratory for Topological Physics, ShanghaiTech University, Shanghai 201210, China}}
\affiliation{State Key Laboratory of Quantum Functional Materials, ShanghaiTech University, Shanghai 201210, China}
\date{\today}
%

%

\begin{abstract}
The microscopic origin of ferromagnetism in correlated materials remains heavily debated, particularly for the competing mechanisms governing insulating versus metallic phases. 
In this work, we theoretically study the electronic structure evolution of CrGeTe$_{3}$ under pressure and provide a consistent explanation to three unique features of this system, i.e. the semiconducting ferromagnetism at low pressure, the metallic ferromagnetism at high pressure, and the enhanced Curie temperature in the metallic phase. 
We propose that it is the reduced electronic correlation and enhanced $d$-$p$ hybridization that universally drive the continuous evolution of CrGeTe$_{3}$ under pressure and glue the three distinct experimental observations.   
Central to our discovery is the dual role of metallicity -- it simultaneously establishes kinetically driven exchange via $d$-$p$ hybridization and enables Stoner-type magnetic instability, with the contribution also from the residual super-exchange. 
Our analyses reveal that {\it intraband} excitations dominate the pressure-enhanced $\omega_p^2$ and $T_c$ correlation. These findings establish $d$-$p$ hybridization and electronic correlation as the bridge between localized and itinerant magnetism, at least, in CrGeTe$_{3}$.
\end{abstract}

\maketitle
\clearpage

\let\oldaddcontentsline\addcontentsline
\renewcommand{\addcontentsline}[3]{}

{\it Introduction --}
The discovery of van der Waals (vdW) magnetic semiconductors, such as CrI$_3$~\cite{mcguire_coupling_2015, CrI3}, CrSiTe$_3$~\cite{lin_ultrathin_2015}, CrGeTe$_{3}$~\cite{gong_discovery_2017}, and Fe$_3$GeTe$_2$~\cite{deiseroth_fe3gete2_nodate, FeGeTe-Yuanbo}, has opened new frontiers in probing correlated electron physics in low-dimensional systems. 
At ambient pressure, CrGeTe$_{3}$ is a ferromagnetic semiconductor with a Curie temperature ($T_c$) of $\sim$ 67 K~\cite{carteaux_crystallographic_1995, 10.1063/1.4822092, PhysRevB.96.054406,PhysRevB.98.214420}, where strong electronic correlations and Cr-Te-Cr superexchange interactions govern its insulating behavior and magnetic order. 
The hybridization between localized Cr-$3d$ orbitals and delocalized Te-$5p$ states introduces pronounced spin-orbit coupling (SOC), stabilizing a perpendicular magnetic anisotropy and, subsequently, a two-dimensional long-range ferromagnetic order~\cite{gong_discovery_2017}.
Moreover, this material exhibits a remarkable pressure-driven transition: under hydrostatic compression, it evolves into a ferromagnetic metal with an enhanced $T_c$ without structural transition~\cite{bhoi_nearly_2021, yu_pressure-induced_2019,ge_raman_2020}.
This abrupt semiconductor-to-metal crossover, accompanied by the persistence and even strengthening of ferromagnetism in the metallic phase, challenges conventional paradigms that often decouple insulating and itinerant magnetism.

According to Raman spectroscopy, a crystalline-to-crystalline phase transition occurs at around 14 GPa, and a crystalline-to-amorphous phase transition occurs above 26.4 GPa~\cite{ge_raman_2020}. Surprisingly, pressure increases Curie temperature of CrGeTe$_3$ from $T_c=67K$ to room temperature~\cite{PhysRevResearch.4.L022040, bhoi_nearly_2021}. 
On one hand, it has been conjectured that increased pressure along with increased bandwidth closes the charge transfer gap between Cr-$d$ and Te-$p$, which dramatically enhances the Curie temperature of CrGeTe$_3$ by super-exchange process~\cite{bhoi_nearly_2021, 10.1063/1.5016568}. 
However, the presence of the local moment required for the super-exchange mechanism is not fully consistent with the itinerancy of Cr-$d$ electrons in the high-pressure metallic phase.
On the other hand, a recent optical conductivity measurement observed a linear increase of plasma frequency square $\omega_p^2$ with pressure $p$ in the metallic state~\cite{PhysRevB.111.L140402}, providing new insight to the metallic ferromagnetism. 
The enhanced $T_c$ and the linear relation with $\omega_p^2$ at high pressure were attributed to a double-exchange mechanism~\cite{DE_PWAnderson_PhysRev.100.675, PhysRevLett.93.147202}, which was earlier proposed for ambient CrGeTe$_{3}$ to account for the shift in X-ray absorption spectroscopy (XAS)~\cite{PhysRevB.101.205125}  and X-ray magnetic circular dichroism (XMCD)~\cite{https://doi.org/10.1002/pssr.202100566}. 
The strong hybridization between Te-$p$ and Cr-$d$ orbitals also favors the contributions from conventional super-exchange. 
A variant of double-exchange was also proposed to explain a similar enhanced $T_{c}$ in electron-doped CrGeTe$_{3}$~\cite{DE_CGT_Wang, DE_CGT_Ivan, CGT_electrondope}. 
Suppressed antiferromagnetic exchange~\cite{CGT_reducedAFM_PhysRevB.111.L020402} and reduced charge-transfer gap~\cite{CGT_reducedGap_PhysRevLett.127.217203} have also been conjectured as possible reasons for the enhanced metallic ferromagnetism. 

While the existence of various interpretations, some experimental details are unfortunately overlooked. 
Specifically, the long-range ferromagnetic order exhibits non-monotonic pressure dependence~\cite{bhoi_nearly_2021, 10.1063/1.5016568, PhysRevB.111.L140402}. 
Below 4 $\sim$ 5 GPa, $T_c$ decreases gradually with increasing pressure before the striking increase in metallic state, which strongly suggests distinct ordering mechanisms dominating the low- and high-pressure regimes. 
The observed $T_c$ minimum near 4$\sim$5~GPa reveals a competitive interplay between these two mechanisms during the pressure-driven crossover. 
Furthermore, this transition occurs at a smaller pressure than that for the semiconductor-metal crossover~\cite{bhoi_nearly_2021}, highlighting the decoupled magnetic ordering and the interaction-induced localization of electrons in this system. 
Thus, the driving force for the semiconductor-metal crossover in the slightly suppressed ferromagnetic background is not yet understood, and  it is not clear whether the super-exchange or double-exchange alone is able to account for the two distinct ferromagnetic responses at low- and hight-pressure. An unified explanation that consistently account for all these phenomena, including also the $T_c$ suppression around the semiconductor-metal crossover, is thus highly desired. 

\begin{figure*}[t]
    \includegraphics[width=1.0\linewidth]{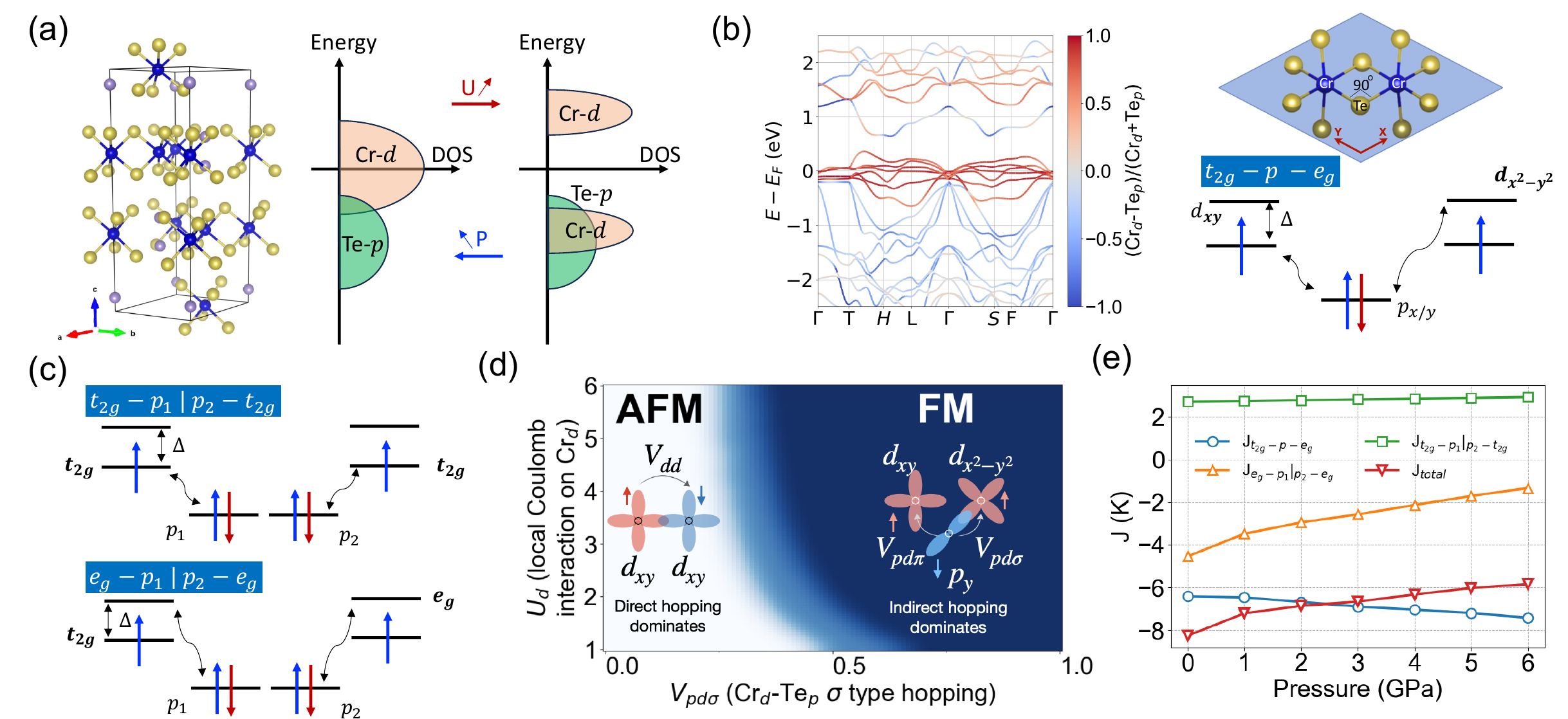}
    \caption{ \label{Fig:ambinent_pressure}
        (a) The conventional cell of CrGeTe$_{3}$ and a schematic diagram of the charge transfer insulating ground state with Cr-$d$ states moving below Te-$p$ states.
        (b) The electronic structure of the nonmagnetic state of CrGeTe$_{3}$ (without SOC) and the local geometry where Cr-Te form three perpendicular planes. 
        In each plane, the angle of Cr-Te-Cr bond is close to 90$^{\circ}$. 
        Due to this specific geometry, only some super-exchange paths are allowed. One example is the $t_{2g}$-$p_x$-$e_{g}$ super-exchange schematically shown in (b).  
        (c) The other two important super-exchange processes are $t_{2g}-p_1|p_2-t_{2g}$ and $e_{g}-p_1|p_2-e_{g}$ super-exchange path.
        (d) The Phase diagram of effective spin couplings as a function of $t_{pd\sigma}$ and $U_d$. Other parameters are fixed as $J_d=0.3U_d$, $J_p=0.1U_p=0.005\ eV$, and $t_{pd\pi}=0.01t_{pd\sigma}$ in this calculation.   
        (e) displays the contribution to the spin coupling from the three super-exchange processes as a function of pressure. The results are evaluated at fixed coulomb repulsion $U_{d0}=4\ eV,\ U_{p0}=0.05\ eV$, and Hund's coupling energy $J_d=0.3U_d,\ J_p=0.1U_p$.
    }
\end{figure*}

In this work, we propose an alternative explanation across different pressures that is consistent with the suppressed insulating ferromagnetism, the semiconductor-metal crossover, and the enhanced metallic ferromagnetism in CrGeTe$_{3}$. 
Our proposal is independent of Cr covalency. 
The reduced electronic correlations, the enhanced \textit{Cr-$d$ and Te-$p$ orbital hybridizations} and the emergent \textit{kinetic ferromagnetic exchange}  are the key observation of this work. 
Crucially, we treat the charge gap and magnetic order as emergent properties of a single electronic landscape without assuming ferromagnetic and semiconducting background in priori, where the interplay between correlation and hybridization evolves continuously under compression, which essentially distinguishes our interpretation from others~\cite{PhysRevB.101.205125, https://doi.org/10.1002/pssr.202100566, DE_CGT_Wang, DE_CGT_Ivan, CGT_electrondope, PhysRevB.111.L140402}.

{\it Semiconducting phase and super-exchange --}
Above $T_{c}$, the short-range ferromagnetic correlation persists in CrGeTe$_{3}$ and the electronic correlation~\cite{PhysRevB.99.161401, PhysRevLett.122.207201} drives it to a charge transfer insulator with Cr-$d$ states moving below Te-$p$ states~\cite{PhysRevLett.123.047203, PhysRevB.98.125127, PhysRevB.101.205125} (see Fig.~\ref{Fig:ambinent_pressure}(a)).  
The presence of local moments at high temperatures is a result of the partially-filled Cr-$d$ shell and the local Hund's coupling. 
The long-range order at low-temperature is stabilized through ferromagnetic super-exchange~\cite{PhysRevLett.123.047203, PhysRevB.101.205125}. 
We follow this argument. While, unlike other theoretical investigations~\cite{PhysRevLett.123.047203, PhysRevB.91.235425, CST_J, PhysRevB.98.125416, D0TC02003F} which determined the effective spin couplings between Cr from density-functional theory (DFT) calculations, 
here we try to extract the nearest-neighbor coupling analytical (see the Supplementary Information), where the leading super-exchange paths become evident and we determine the condition for decreasing $T_c$ in the insulating states.  

The ferromagnetic super-exchange is mainly contributed by the effective coupling between the $t_{2g}$ and $e_{g}$ states of the two neighboring Cr atoms~\cite{PhysRevLett.123.047203, PhysRevB.101.205125}. 
Due to symmetry constraints, only the $t_{2g}$ states  are allowed to couple with $e_{g}$ states at neighboring Cr through hybridization with the Te-$p$ states. 
This exchange process is denoted as $t_{2g}-p-e_{g}$ in Fig.~\ref{Fig:ambinent_pressure}(b), where, as an example, Cr-$d_{xy}$ electron couple with the neighboring Cr-$d_{x^2-y^2}$ state via the intermediate Te-$p_{x/y}$ states. 
 We employ the second-order perturbation theory to derive the effective spin couplings~\cite{KANAMORI195987, Weihe_1997}. 
Relevant model parameters are extracted from the nonmagnetic electronic structure, including crystal field $\Delta$ between Cr-$t_{2g}$ states and Cr-$e_{g}$ states and the hopping amplitude parametrized in Slater-Koster integral~\cite{PhysRev.94.1498}. 
In addition to the super-exchange $t_{2g}-p-e_{g}$, two neighboring $t_{2g}$ states can also couple. But such a process involves two different Te-$p$ electrons and it favors an antiferromagnetic exchange.  
Two different Te-$p$ electrons can also couple two neighboring $e_{g}$ states as illustrated in Fig.~\ref{Fig:ambinent_pressure}(c), and this corresponds to an effective ferromagnetic exchange. 
In Fig.~\ref{Fig:ambinent_pressure}(c), the latter two processes are denoted as $t_{2g}-p_{1}|p_{2}-t_{2g}$ and $e_{g}-p_{1}|p_{2}-e_{g}$, respectively. 

After collecting contributions from all three exchange processes (see the Supplementary Information for more details), we obtained the effective spin coupling of the neighboring Cr atoms as shown in Fig.~\ref{Fig:ambinent_pressure}(d). 
The Cr-$d$ and Te-$p$ $\sigma$-type coupling strength $V_{pd\sigma}$ and the local electronic correlation $U_d$ of Cr-$d$ electrons are taken as free parameters. 
The choice of $V_{pd\sigma}$ is consistent with the estimation from wannier projection. 
In the phase shown in white the spin coupling is antiferromagnetic (AFM). In phase colored in blue the spin coupling is ferromagnetic and the coupling strength decreases with the increase of pressure. In the crossover regime between the AFM and FM, the spin coupling is ferromagnetic but it increases with pressure. 
For realistic parameters relevant to CrGeTe$_{3}$, i.e., $V_{pd\sigma}\sim 1$ eV and $U_{d}\sim 4.0$ eV, the resulting spin coupling is ferromagnetic and decreases with pressure in semiconducting phase ($p \le 4$ GPa), which is highly consistent with experiment.

\begin{figure}[t]
\centering
\includegraphics[width=\linewidth]{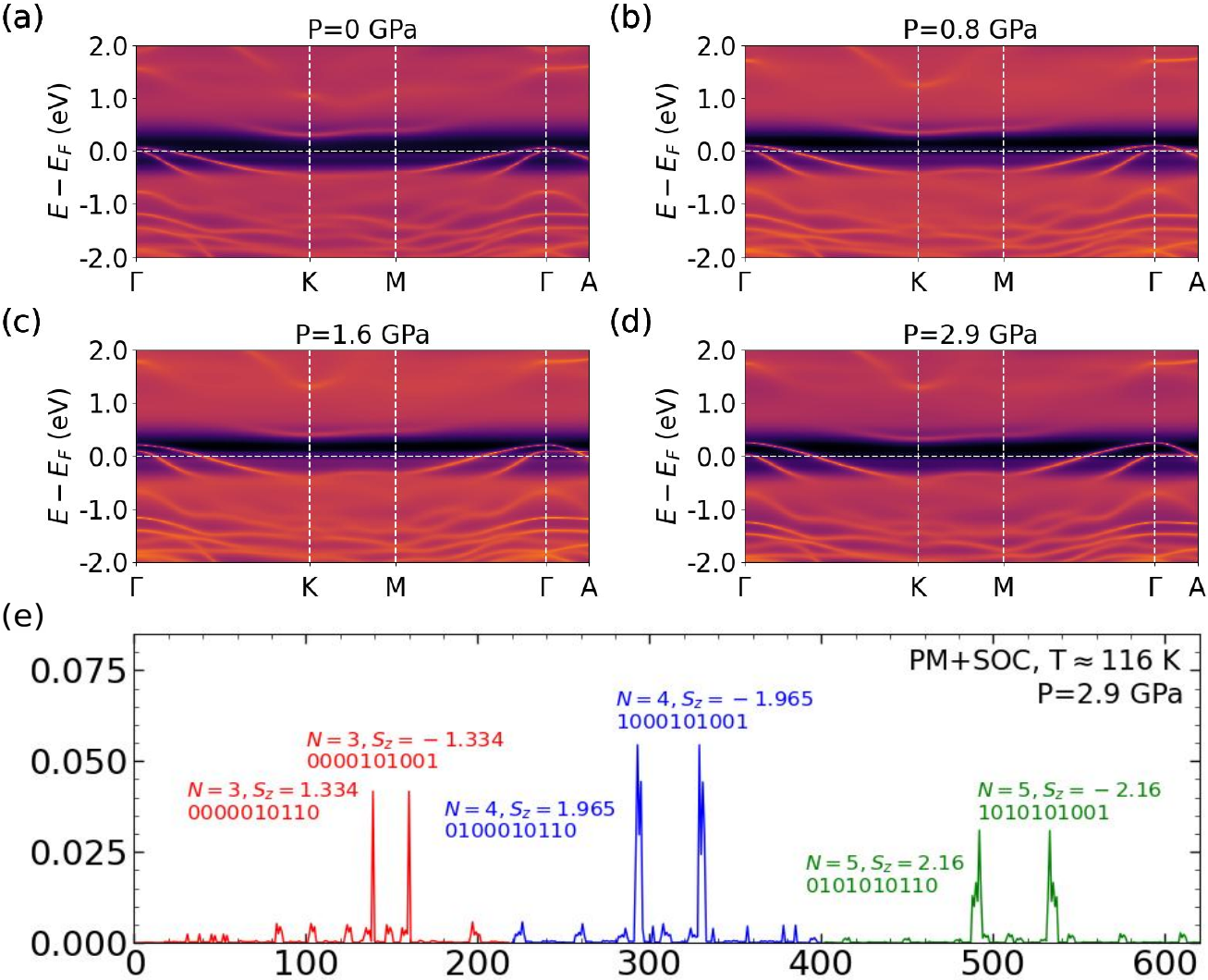}
\caption{Evolution of CrGeTe$_{3}$ charge gap under different pressures. (a)-(d) corresponds to the spectral functions at ambient pressure, $P = 0.8$ GPa, 1.6 GPa, and 2.9 GPa, respectively. (e) The contribution of the many-body atomic states to the ground state at $T\approx 116$ K and $P = 2.9$ GPa with paramagnetic condition. The presence of local moments are characterized by the three pairs of states in $N=3, 4$, and $5$ subspaces.}
\label{Fig:DMFT}
\end{figure} 
{\it Semiconductor-metal crossover --} 
After understanding the ferromagnetic semiconducting phase at low pressure, it is straightforward to understand the semiconductor-metal crossover under external pressures. 
Generally, pressures modify electronic structures in two complementary ways. 
On the one hand, pressures change the crystal structure by adjusting the atomic spacings, leading to volume or symmetry changes~\cite{CrSiTe3-pressure, PhysRevB.102.144525}.
On the other hand,  compressing crystals often leads to a stronger overlap of the wavefunctions, enhancing the electron kinetic energy.  
Consequently, the relative strength of Coulomb energy against kinetic energy is reduced (see Tab. S1 in the Supplementary Information for $U$ values estimated from constraint random-phase approximation). 
Thus, increasing pressure often leads to the delocalization of electrons in the Mott and charge-transfer insulators. 

Figure~\ref{Fig:DMFT} displays the spectral function of CrGeTe$_{3}$ at different pressures calculated with DFT + dynamical mean-field theory (DMFT) implemented in eDMFT package~\cite{Haule2010} using a hybridization expansion continuous-time quantum Monte Carlo method~\cite{PhysRevLett.97.076405, Haule2007, Gull_2008} as impurity solver.
$U_d=4.0$ and $J_d = 0.8$ eV were taken as impurity Coulomb parameters in all calculations.  
The semiconductor-metal crossover can be seen from the evolution of spectral function with pressure. 
Below 2.9 GPa, strong electronic correlations suppress spectral weight at the Fermi energy, with a clear optical gap observed consistent with experiment~\cite{PhysRevB.111.L140402}.
In the ferromagnetic ordered phase, the remaining two bands at the Fermi level will also be removed. 
Since identical $U_d$ and $J_d$ values are maintained across pressures, the crossover primarily stems from lattice compression and increased wavefunction overlap, which enhance kinetic energy relative to correlations. Additional metallicity could emerge with further $U_d$ reduction, consistent with prior studies~\cite{PhysRevLett.123.047203, PhysRevB.108.125142}. 
Notably, local moments persist at temperature significantly exceeding $T_{c}$ under pressure. 
As shown in Fig.~\ref{Fig:DMFT}(e) ($T\sim 116$ K), local moments contribute in pairs with the overall magnetization being zero.
The ground state is dominantly contributed by atomic states in $N=3, 4$, and 5 subspace with notable local moments.    
Further evidence of persisting local moments in the metallic phase is shown in Fig. S6 of the Supplementary Information, highlighting the persistent super-exchange contribution even in the metallic phase.
A persistent local moment is also required by the double-exchange~\cite{DE_PWAnderson_PhysRev.100.675, PhysRevLett.93.147202} and ferromagnetic kinetic-exchange~\cite{PhysRevResearch.2.033430} mechanisms. 

{\it Mechanism for metallic ferromagnetism --}
After the charge gap is closed, the magnetic moments in the semiconducting states gradually become delocalized. 
Although the super-exchange mechanism still contributes in the metallic states, it is not sufficient to explain the astonishing enhancement of $T_c$ at high pressure. A new mechanism for metallic ferromagnetism is highly demanded. 
One candidate mechanism is the coexistence of localized and delocalized $d$ electrons, which are in charge of the ferromagnetic super-exchange and metallicity independently~\cite{PhysRevB.70.052404, Dong_2022, PhysRevB.101.205125}. 
Additionally, we propose that the joint operation of an enhanced $d$-$p$ hybridization and the local Hund's coupling in CrGeTe$_{3}$ make the ferromagnetic kinetic-exchange the leading driving force for the metallic ferromagnetic couplings in CrGeTe$_{3}$~\cite{bhoi_nearly_2021}.  

\begin{figure}[t]
    \includegraphics[width=1.0\linewidth]{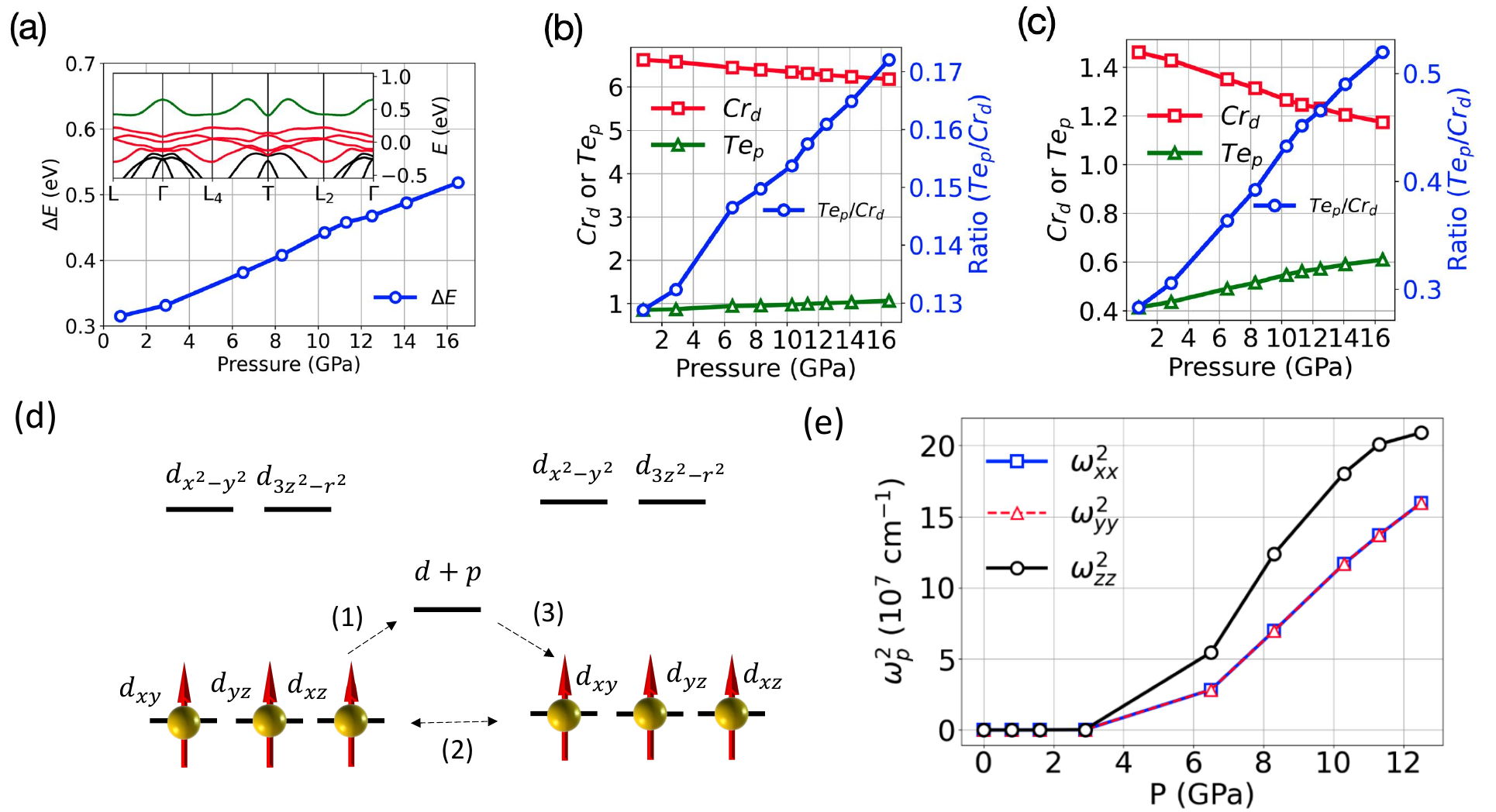}
    \caption{ \label{Fig:d-p_hybrid}
        (a) With the increase of pressure, one band gradually separates from the other $t_{2g}$ bands around the Fermi level, promoting the kinetic ferromagnetic exchange. The energy difference between this band (shown as green in the inset plot) and the rest four  bands (shown as red in the inset plot) increases with pressure. 
        (b) and (c) monitor the relative contributions of  Cr-$d$ electrons and Te-$p$ electrons to (c) the green band and to (d) the four red bands, respectively. 
        (d) shows a schematic diagram of the emerging kinetic ferromagnetic exchange due to the enhanced  $d$-$p$ hybridization, which requires only three single-particle hoppings. 
        (e)  \textit{ab initio} evaluation of plasma frequencies square at different pressures from intraband excitations. 
    }
\end{figure}

The inset of Fig.~\ref{Fig:d-p_hybrid}(a) displays the DFT electronic structure of CrGeTe$_{3}$ at $P=10.3$ GPa. 
We found that one band from the $t_{2g}$ subset at the Fermi level gradually separates from the others and moves to a slightly higher binding energy upon increasing pressure. This band is highlighted in green in Fig.~\ref{Fig:d-p_hybrid}(a), and the rest of $t_{2g}$ bands are colored in red. Their average energy difference is shown as the blue line with an empty circle, which monotonically increases with pressure. 
We note that the orbital characters of these bands are no longer $t_{2g}$ in the strict sense as the $d$-$p$ hybridization becomes much stronger in the high-pressure phase. 
As evident in Fig.~\ref{Fig:d-p_hybrid} (b),  the Cr-$d$ states dominate the atomic contributions to the Bloch wave function of the red band in Fig.~\ref{Fig:d-p_hybrid}(a). In this case, Te-$p$ orbitals contribute very little to these bands. 
While for the band in green, Te-$p$ orbital contribution becomes much more pronounced (see Fig.~\ref{Fig:d-p_hybrid}(c)), offering a new channel for exchanging equal-spin electrons between neighboring Cr atoms. 
We note that the $d$-$p$ hybridization discussed here is different to the strong covalent binding observed at ambient pressure~\cite{PhysRevB.101.205125}, which  formed by the Te-$p$ and Cr $e_g$ orbitals. 
The kinetic ferromagnetic exchange at high pressure benefits from the enhanced hybridization of Te-$p$ with the Cr-$t_{2g}$ orbitals, which makes the exchange of spins much easier. 

Figure~\ref{Fig:d-p_hybrid}(d) depicts the electrons with equal spin traveling between two Cr $t_{2g}$ states via the emergent Te-$p$ state with a slightly higher energy.  
Electrons from the left Cr state move to the unoccupied Te-$p$ states and leave a hole behind, which will be filled by one Cr-$d$ electron from the neighboring right site. 
The unstable electron in the Te-$p$ states then moves to the right Cr $t_{2g}$ states. 
The local Hund's coupling at each Cr site prefers all electron spins to align parallelly, leading to the unusual synergy of metallicity and long-range ferromagnetic order.  
Compared to the super-exchange process at low pressure, the ferromagnetic kinetic exchange~\cite{PhysRevResearch.2.033430} takes only three steps to connect two neighboring Cr states instead of four required by conventional super-exchange. 
Thus, with the enhanced $d$-$p$ hybridization and remnant local Hund's coupling, the ferromagnetic order can be more efficiently established than super-exchange.
Compared to the double-exchange mechanism~\cite{PhysRevB.111.L140402}, no Cr covalency is required. 
The change from one mechanism at low-pressure to the other mechanism at high-pressure is also more natural for kinetic-exchange than for the double-exchange, as the former develops smoothly with the continuous change of electronic structure.
We note that our ferromagnetic kinetic-exchange scenario is similar to the double-exchange mechanism proposed for the electron-doped CrGeTe$_{3}$ where a boost of $T_c$ is also observed~\cite{DE_CGT_Wang}.  
In our scenario, the states participating the spin-exchange are not the high-energy $e_g$ states but the additional $d-p$ hybridized state separating from $t_{2g}$, which stay below the $e_g$ states and, thus, favor a stronger exchange amplitude than the double-exchange~\cite{PhysRevResearch.2.033430}. 

{\it Enhanced ferromagnetism under pressure --}
In this section, we provide further evidence to support the enhanced metallicity and $d$-$p$ hybridization as the driving force for the boosted $T_c$ in the high-pressure phase. 
Optical measurements reveal a linear scaling between $T_c$ and the squared plasma frequency $\omega_p^2$ above 4~GPa~\cite{PhysRevB.111.L140402}.
This empirical relationship originates from the pressure-induced semiconductor-to-metal crossover, which simultaneously enhances carrier density $n$ through metallization and reduces effective mass $m^*$ via weakened electronic correlations. According to the Drude formalism $\omega_p^2 = 4\pi e^2 n\hbar^2/m^*$, these cooperative effects produce the observed $\omega_p^2$ enhancement.

Our first-principles calculations quantitatively track this evolution in Fig.~\ref{Fig:d-p_hybrid}(e). 
Below 4~GPa, CrGeTe$_3$ maintains its semiconducting nature with negligible Fermi-level carriers, rendering $\omega_p^2$ effectively zero. The semiconductor-metal crossover activates Cr-$d$ and Te-$p$ hybridized states at $E_F$, initiating {\it intraband} excitations that generate finite Drude weight. Above the critical pressure, coherent metallic transport emerges through dominant intraband contributions to $\omega_p^2$, which increases quickly with pressure approximately in a linear form. 

The remarkable consistency between our computational results in Fig.~\ref{Fig:d-p_hybrid}(e) and optical measurements of $\omega_p^2$~\cite{PhysRevB.111.L140402} underscores the critical influence of metallization-induced Cr-$t_{2g}$ and Te-$p$ hybridization at the Fermi level. 
The enhanced metallicity aligns qualitatively with Stoner's criterion for ferromagnetic instability, which suggests a hybrid mechanism where \textit{coexisting} local moments and itinerant carriers cooperate to amplify $T_c$. 
The intraband excitation and the associated kinetic ferromagnetic exchange proposed in this work are consisetnt with most of the available experiment observations, including also the doping~\cite{https://doi.org/10.1002/aelm.202300609} enhanced effect. 
Crucially, the pressure-tuned lattice contraction strengthens both ingredients concurrently -- a cooperative effect that transcends the conventional classification of magnetism into purely localized or itinerant categories.

\begin{figure}[t]
    \includegraphics[width=\linewidth]{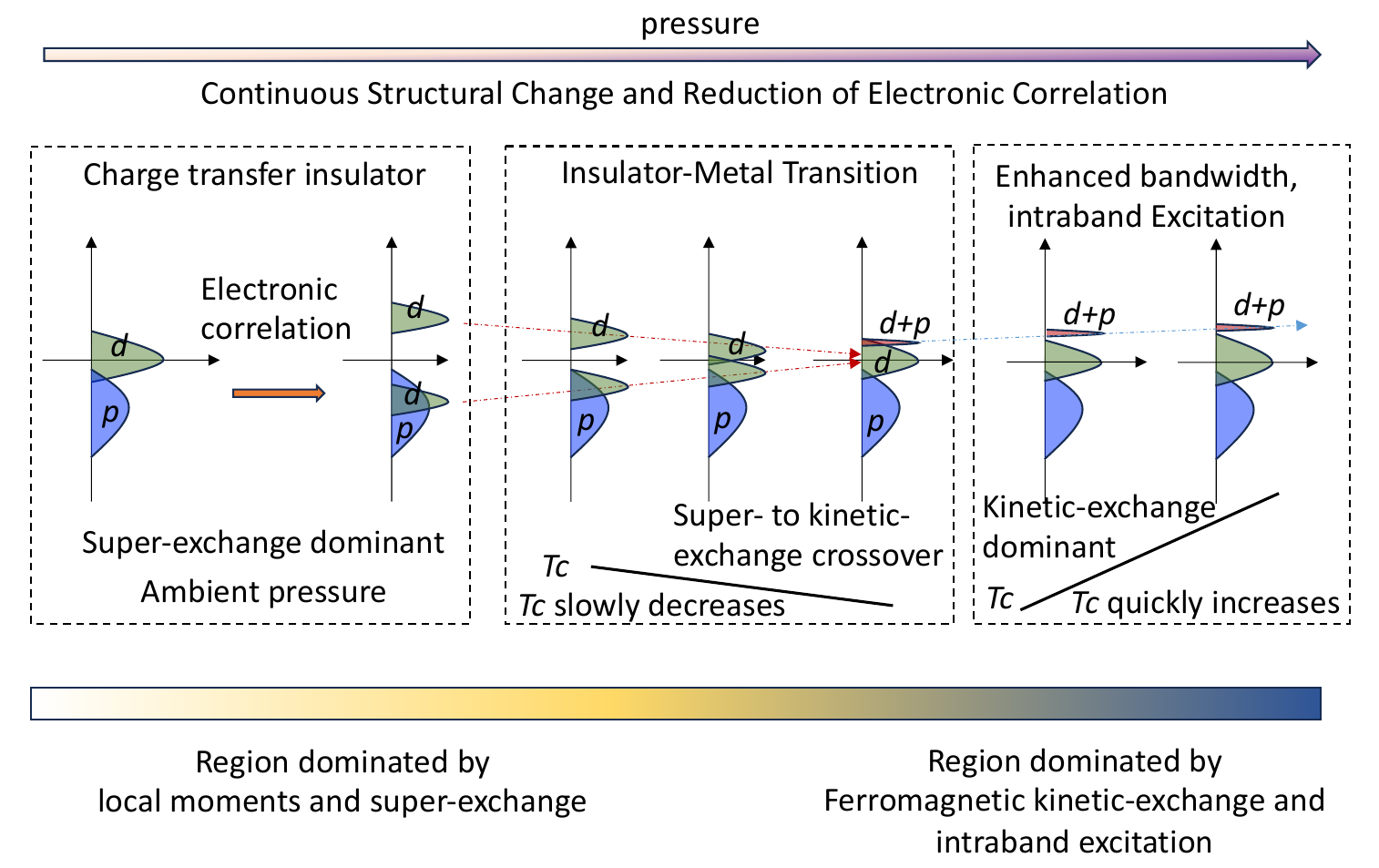}
    \caption{
       Schematic summary of the evolution from super-exchange to kinetic-exchange under the  continuous evolution of electronic structure with pressure. 
    }
\label{Fig:summary}
\end{figure}
{\it Discussions and conclusions --}
In this work, we theoretically studied the pressure-induced semiconductor-metal crossover and the enhanced metallic ferromagnetism in CrGeTe$_{3}$. 
Our proposal is summarized in Fig.~\ref{Fig:summary}. 
We showed that the semiconductor-metal crossover emerges as a synergy of structural compression and diminished electronic correlations, where lattice contraction optimizes $d$-$p$ orbital overlap while pressure-induced bandwidth enhancement suppresses correlation effects. 
The peculiar electronic structure under pressure highlights an emerging $d-p$ hybridized band offering a new channel for ferromagnetic kinetic-exchange.
As pressure increases, the $d-p$ hybridization is enhanced but the electronic correlations are reduced. As a result, the effective ferromagnetic exchange coupling quickly increases~\cite{PhysRevResearch.2.033430} consistent with the enhanced $T_c$ in the metallic states~\cite{PhysRevB.111.L140402}. 
Together with the local Hund's coupling, this provides a new mechanism for ferromagnetism enabling itinerant electrons to participate in ordering. 
Crucially, the pressure-enhanced $d$-$p$ orbital hybridization establishes parallel conduction channels that amplify $n$ and release $m^*$, enhancing $\omega_p^2$ as well as the conventional Stoner instability. 
These findings carry profound implications. 
The correlation of high $T_c$ with the \textit{intraband} excitations provides \textit{compelling evidence} for metallicity as one important gradient for magnetic ordering. 
Moreover, the scaling of $T_c$ with $\omega_p^2$ challenges purely localized or itinerant magnetism doctrines, instead demanding a hybrid exchange paradigm accounting for both momentum-resolved spin dynamics and real-space correlation residues. 
We believe that the enhanced $T_c$ benefits from the joint contribution of ferromagnetic kinetic-exchange, the residual super-exchange, and the Stoner instability. 

\begin{acknowledgments}
This work was supported by the National Key R\&D Program of China (No. 2022YFA1402703), Sino-German Mobility program (No. M-0006), and Shanghai 2021- Fundamental Research Area (No. 21JC1404700). Part of the calculations was performed at the HPC Platform of ShanghaiTech University Library and Information Services, and the School of Physical Science and Technology.
\end{acknowledgments}

\bibliographystyle{apsrev4-1}
\bibliography{MyPaperRef.bib}

\clearpage
\onecolumngrid

\renewcommand{\thefigure}{S\arabic{figure}}
\renewcommand{\thetable}{S\arabic{table}}
\renewcommand{\theequation}{S\arabic{equation}}
\setcounter{figure}{0} 
\setcounter{table}{0}
\setcounter{equation}{0}

\makeatletter
\def\@hangfrom@section#1#2#3{\@hangfrom{#1#2}#3}
\def\@hangfroms@section#1#2{#1#2}
\makeatother

\begin{center}
{\large\it Supplementary Information}\\
\vspace{0.5cm}

{\large\bf Emerging kinetic-exchange for the enhanced metallic ferromagnetism in CrGeTe$_3$ under pressure}

\vspace{0.3cm}

Jiaming Liu$^{1}$,
Xuefeng Zhang$^{1}$, 
Hongbin Qu$^{1}$, 
Xiaoqun Wang$^{2}$,
Hai-Qing Lin$^{2}$,
and Gang Li$^{1, 3, 4, *}$\\
\vspace{0.2cm}
${}^1${\small\it School of Physical Science and Technology, ShanghaiTech University, Shanghai 201210, China}\\
${}^2${\small\it School of Physics, Zhejiang University, Hangzhou 310027, Zhejiang, China} \\
${}^3${\small\it School of Physical Science and Technology, ShanghaiTech University, Shanghai 201210, China}\\
${}^3${\small\it State Key Laboratory of Quantum Functional Materials, ShanghaiTech University, Shanghai 201210, China}
\end{center}

\vspace{0.2cm}

\let\addcontentsline\oldaddcontentsline
\pagenumbering{arabic}
\setcounter{page}{1}

\tableofcontents 

\section{Details of {\it ab-initio} calculations}
To obtain the effective exchange intensity of CrGeTe$_3$, we first construct the paramagnetic effective Hamiltonian under different pressures. We employ the Vienna Ab initio Simulation Package (VASP) for the first-principles calculations~\cite{noauthor_phys_nodate,kresse_efficiency_1996} and WANNIER90 package for wannier projection~\cite{noauthor_phys_nodate-1,koepernik_symmetry-conserving_2023} of the tight-binding model. 
Structure was relaxed at atmospheric and high pressures until the forced is less than 0.02 eV/Å with plane-wave cut-off energy of 400 eV. 
To study the interplay of electron correlation and magnetism in Cr$_2$Ge$_2$Te$_6$, we performed the all-electron charge self-consistent DFT+DMFT calculations by using the embedded-DMFT package~\cite{Haule2010}.
The DFT calculations in DFT+DMFT were done with the full-potential augmented plane-wave + local orbitals (APW+lo) scheme as implemented in WIEN2k~\cite{Blaha2020}.
We used GGA-PBE functional and set RKmax = 7.0 and use a $10\times10\times$ 10 $k$-mesh for the BZ sampling. In DMFT calculations, we set the on-site Coulomb interaction as $U$ = 4 eV and the Hund's exchange interaction $J$ = 0.8 eV for Cr-3$d$ orbitals.
The quantum impurity problem was solved by using the continuous-time quantum Monte Carlo (CTQMC) method~\cite{Werner2006,Werner2006a,Haule2007,Gull2011}.
SOC was included self-consistently in our PM DMFT calculation. 

\section{Calcualtion of super-exchange interactions}
In this work, we followed the Goodenough-Kanamori-Anderson (GKA) rules for super-exchange interactions~\cite{khomskii_transition_2014} to obtain the effective couplings $J$ between two neighbor Cr cations at small pressure regime. 
In the semiconducting phase, due to the strong electronic correlations, the Cr-$d$ states and the associated magnetic moments become fully localized. 
Thus, the magnetic coupling between neighboring Cr atoms is mainly contributed by super-exchange. 
Starting from the paramagnetic high-temperature phase, as guided by the DFT calculation, we depicted the relative energy positions of Cr-$t_{2g}$, Cr-$e_g$, and Te-$p$ in Fig.~\ref{Fig:ambinent_pressure}(b). 
There are three super-exchange possibilities that connect two neighboring Cr atoms within finite number of single-electron hopping. 
They are shown in Fig.~\ref{Fig:ambinent_pressure}(b) and (c). 
We consider the following Hamiltonian which accounts for the energy difference of Cr-$t_{2g}$, Cr-$e_{g}$, Te-$p$ states, as well as the local Coulomb interactions $U_{d}$, $U_{p}$, Hund's coupling $J_d$, $J_p$. 
\begin{align}\label{EqS:local_H}
		H&=\sum_{\alpha,\sigma} \epsilon_d^{\alpha}(n_{\sigma}^{a,\alpha} + n_{\sigma}^{b,\alpha})  + \epsilon_p\sum_{\gamma,\sigma}n_{\sigma}^{\gamma} \nonumber 
		+\sum_{\alpha\beta,\sigma}(t_{ab}^{\alpha\beta}\hat{a}_{\alpha\sigma}^{\dagger}\hat{b}_{\beta\sigma}+h.c.) 
		+\sum_{\alpha\gamma,\sigma}(t_{ap}^{\alpha\gamma}\hat{a}_{\alpha\sigma}^{\dagger}\hat{p}_{\gamma\sigma}+t_{bp}^{\beta\gamma}\hat{b}_{\beta\sigma}^{\dagger}\hat{p}_{\gamma\sigma}+h.c.) \\
		&+U_d\sum_{\alpha}n_{\uparrow}^{\alpha}n_{\downarrow}^{\alpha} +(U_d-2J_d)\sum_{\alpha\neq\beta}n_{\uparrow}^{\alpha}n_{\downarrow}^{\beta} 
		+(U_d-3J_d)\sum_{\alpha<\beta\sigma}n_{\sigma}^{\alpha}n_{\sigma}^{\beta} \nonumber \\
		&+U_p\sum_{\gamma}n_{\uparrow}^{\gamma}n_{\downarrow}^{\gamma} + (U_p-2J_p)\sum_{\gamma\neq\delta}n_{\uparrow}^{\gamma}n_{\downarrow}^{\delta}
		+(U_p-3J_p)\sum_{\gamma<\delta\sigma}n_{\sigma}^{\gamma}n_{\sigma}^{\delta}\;,
\end{align}
where $\alpha$ and $\beta$ label the five Cr-$d$ states. $\gamma$ and $\delta$ correspond to the orbital indices of the three Te-$p$ states. 
For clarity, the creation/annihilation operators for the left and the right Cr atoms are denoted as $\hat{a}^\dagger/\hat{a}$ and $\hat{b}^\dagger/\hat{b}$.
We further assume that all three Cr-$t_{2g}$ states are degenerate and their energy level is $\epsilon^d$. 
The Cr-$e_{g}$ states are also degenerate but with a higher energy level $\epsilon_d+\Delta$.
Due to the symmetry of wavefunctions, the single-electron hopping $t_{ab}^{\alpha\beta}$, $t_{ap}^{\alpha\gamma}$ depend on the orientation of hopping and can be different for the left and the right Cr-$d$ electrons. 
One simple parameterization is to adopt the Slater-Koster parameters to account for the directional dependence of these hopping. 

To obtain the parameters in Eq.~(\ref{EqS:local_H}), we downfold the {\it ab-initio} electronic structure to a tight-binding model by using Wannier90 and extract the local Hamiltonian that contains all the single-particle parameters, i.e.  $\epsilon_d$, $\Delta$, $t_{ab}^{\alpha\beta}$, $t_{ap}^{\alpha\gamma}$, and $t_{ap}^{\alpha\gamma}$. 
Meanwhile, $U^d$, $J^d$, $U^p$, and $J^p$ are considered as free parameters.  

\subsection{Super-exchange process $t_{2g}-p-e_{g}$}
\begin{figure}[htbp]
\centering
\includegraphics[width=\linewidth]{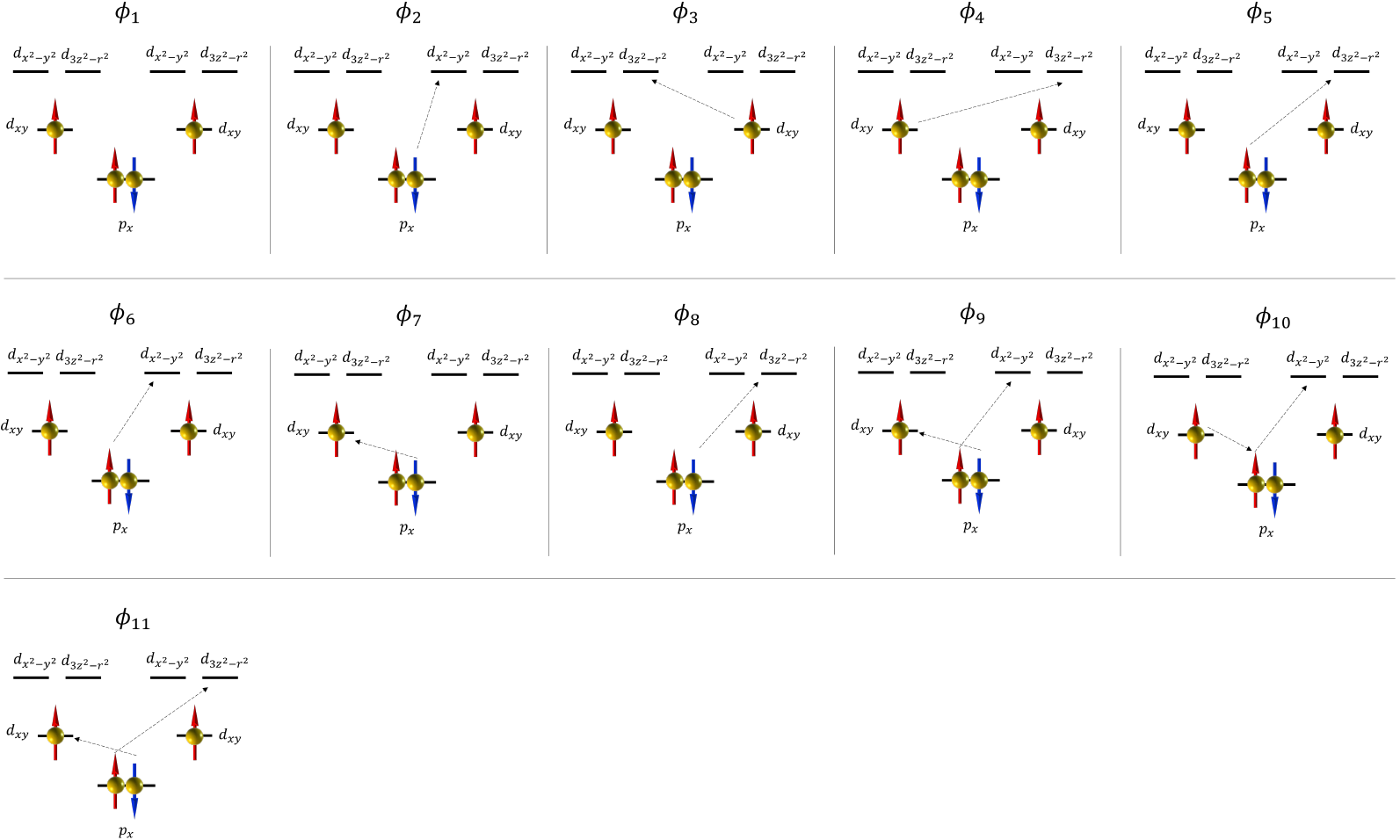}
\caption{Exchange processes considered for the correction to state $\phi_1$ with parallel-spin configuration.  In addition to the excited states that differ from $\phi_1$ by one-step of electron-hopping, the excitation of both electrons at $p_{x}$ state is also considered.}
\label{FigS:parallel_1}
\end{figure}

This super-exchange involves the same Te-$p$ electron as media to connect two neighboring Cr atoms. Take one of the Cr-O-Cr planes shown in Fig.~\ref{Fig:ambinent_pressure}(b) as example, 
it involves $d_{xy}$, $d_{x^2-y^2}$, $d_{3z^2-r^2}$ from both Cr atoms and $p_{x}$, $p_{y}$ from the ligand Te atom. 
We take the following basis functions in the occupation number basis as our starting point. 
\begin{align}\label{EqS:state}
|\phi\rangle = |a_{xy},a_{x^2-y^2}, a_{3z^2-r^2}, p_{x}, b_{xy}, b_{x^2-y^2}, b_{3z^2-r^2}\rangle \;,
\end{align}
where $a$ and $b$ correspond to the $d$ states at the left and right Cr atoms. Each state can be occupied by either up- or down-spin electron. 
According to the DFT electronic structure, $t_{2g}$ states are singly occupied, $e_{g}$ states are empty, and the Te-$p$ states are fully occupied. Consider a ferromagnetic configuration of the two neighboring Cr atoms, we have the ground state as  
$\phi_1 = |\uparrow, 0, 0\rangle\otimes|\uparrow\downarrow\rangle\otimes|\uparrow, 0, 0\rangle$, which corresponds to a state with two singly occupied $d_{xy}$ states, a doubly occupied $p_x$ state. 
The Hamiltonian matrix element $\epsilon_1 = \langle\phi_1|H|\phi_1\rangle = 2\epsilon_d + 2\epsilon_p + U_p$. 
Due to the energy difference and Coulomb repulsion, single-electron hopping brings it to other states with higher energy, serving as the high-energy sector that will be removed to obtain the effective Hamiltonian of $\phi_1$.
As there are 3 up-spin and 1 down-spin electrons to fill 7 up-spin states and 7 down-spin states, in total there will be $C_{7}^3 \times C_{7}^1 = 245$ states consisting the Hilbert space. 
Due to practical reasons, it is not feasible to include all 245 states in the super-exchange calculation. 
One can classify these states according to the single-particle hopping steps they required to transfer to $\phi_1$.   
In this work, we restrict our calculations mainly to those states that only differ from $\phi_1$ by one step of single-electron hopping. There are 7 such states, as shown below. 
\begin{align}
\begin{array}{lll}
\phi_{2} = |\uparrow, 0, 0 \rangle\otimes| \uparrow \rangle\otimes| \uparrow, \downarrow, 0\rangle\;,  & \epsilon_{2} =  3\epsilon_d + \Delta + \epsilon_p + (U_d - 2J_d)\;,  & p_{x,\downarrow} \rightarrow \text{ right } d_{x^2-y^2,\downarrow}\;, \nonumber \\
\phi_{3} = |\uparrow,0,\uparrow\rangle\otimes|\uparrow\downarrow\rangle\otimes |0,0,0\rangle\;, & \epsilon_3 = 2\epsilon_d + \Delta + 2\epsilon_p + (U_d - 3J_d) + U_p\;, & \text{ right } d_{xy, \uparrow} \rightarrow \text{ left } d_{3z^2-r^2}\;, \cr
\phi_{4} = |0,0,0\rangle\otimes|\uparrow\downarrow\rangle\otimes |\uparrow,0,\uparrow \rangle\;,  & \epsilon_4 = 2\epsilon_d + \Delta + 2\epsilon_p + (U_d - 3J_d) + U_p\;, & \text{ left } d_{xy, \uparrow} \rightarrow \text{ right } d_{3z^2-r^2}\;, \cr
\phi_{5} = |\uparrow, 0, 0 \rangle\otimes| \downarrow \rangle\otimes| \uparrow, 0, \uparrow\rangle\;,  & \epsilon_5 =  3\epsilon_d + \Delta + \epsilon_p + (U_d - 3J_d)\;,      & p_{x,\uparrow} \rightarrow \text{ right } d_{3z^2-r^2,\uparrow}\;, \nonumber \\
\phi_{6} = |\uparrow, 0, 0 \rangle\otimes| \downarrow \rangle\otimes| \uparrow, \uparrow, 0\rangle\;,  & \epsilon_{6} =  3\epsilon_d + \Delta + \epsilon_p + (U_d - 3J_d)\;,      & p_{x,\uparrow} \rightarrow \text{ right } d_{x^2-y^2,\uparrow}\;, \nonumber \\
\phi_{7} = |\uparrow\downarrow, 0, 0 \rangle\otimes| \uparrow \rangle\otimes| \uparrow, 0, 0\rangle\;, & \epsilon_7 =  3\epsilon_d + \epsilon_p + U_d\;,     & p_{x,\downarrow} \rightarrow \text{ left } d_{xy,\downarrow}\;, \nonumber \\
\phi_{8} = |\uparrow, 0, 0 \rangle\otimes| \uparrow  \rangle\otimes| \uparrow, 0, \downarrow\rangle\;, & \epsilon_{8} =  3\epsilon_d + \Delta + \epsilon_p + (U_d - 2J_d)\;,  & p_{x,\downarrow} \rightarrow \text{ right } d_{3z^2-r^2,\downarrow}\;.
\end{array}
\end{align}
We note that electron from $p_x$ state cannot move to the left $d_{x^2-y^2}$, $d_{3z^2-r^2}$ states, and to the right $d_{xy}$ state by just one step electron hopping owing to the orthogonalization of their wavefunctions (cf. the Slater-Koster integrals).
Only the direct hopping between the two Cr atoms can connect them by one-step hopping, so at least two steps are required for super-exchange. 
We further include the hopping processes of both Te-$p$ electrons to Cr atoms. There are two more states.   
\begin{align}
\phi_9       &= |\uparrow\downarrow, 0, 0\rangle\otimes|0\rangle \otimes |\uparrow, \uparrow, 0\rangle\;, \epsilon_9 = 4\epsilon_d + \Delta +  2U_d - 3J_d\;, p_{x,\downarrow}\rightarrow \text{ left } d_{xy,\downarrow} \text { and } p_{x\uparrow} \rightarrow \text{ right } d_{x^2-y^2}\;. \nonumber \\
\phi_{10}       &= |0, 0, 0\rangle\otimes|\uparrow\downarrow\rangle \otimes |\uparrow, \uparrow, 0\rangle\;, \epsilon_{10} = 2\epsilon_d + \Delta + 2\epsilon_p + U_d - 3J_d + U_p\;, p_{x,\uparrow}\rightarrow \text{ right } d_{x^2-y^2} \text { and } \text{ left } d_{xy,\uparrow} \rightarrow p_{x}\;. \nonumber \\
\phi_{11}  &=  |\uparrow\downarrow, 0, 0\rangle\otimes|0\rangle \otimes |\uparrow, 0, \uparrow\rangle\;, \epsilon_{10} = 4\epsilon_d + \Delta +  2U_d - 3J_d\;, p_{x,\downarrow}\rightarrow \text{ left } d_{xy,\downarrow} \text { and } p_{x\uparrow} \rightarrow \text{ right } d_{3z^2-r^2}\;.
\end{align}
Figure~\ref{FigS:parallel_1} displays these states and how they are connected to $\phi_1$ by single-particle hopping. 

With the 10 states as the basis, one can write down the matrix form of Hamiltonian Eq.~(\ref{EqS:local_H}) as
\begin{align}
\begin{pmatrix}
\epsilon_1 & \frac{\sqrt{3}}{2}V_{pd\sigma} & \frac{\sqrt{3}}{4}(V_{dd\sigma} -V_{dd\delta}) & \frac{\sqrt{3}}{4}(V_{dd\sigma} -V_{dd\delta}) & -\frac{1}{2}V_{pd\sigma} & \frac{\sqrt{3}}{2}V_{pd\sigma} & V_{pd\pi}  & -\frac{1}{2}V_{pd\sigma} \cr
 \frac{\sqrt{3}}{2}V_{pd\sigma} & \epsilon_2 & 0 & 0 & 0 & 0 & 0 & 0 \cr
 \frac{\sqrt{3}}{4}(V_{dd\sigma} -V_{dd\delta}) & 0 & \epsilon_3 & 0 & 0 & 0 & 0 & 0 \cr
 \frac{\sqrt{3}}{4}(V_{dd\sigma} -V_{dd\delta}) & 0 & 0 & \epsilon_4 & V_{pd\pi} & 0 & 0 & 0  \cr
 -\frac{1}{2}V_{pd\sigma} & 0 & 0 & V_{pd\pi} & \tilde{\epsilon}_5 & 0 & V_{57} & 0 \cr
  \frac{\sqrt{3}}{2}V_{pd\sigma} & 0 & 0 & 0 & 0 & \tilde{\epsilon}_6 & V_{67} & 0 \cr
 V_{pd\pi} & 0 & 0 & 0 & V_{75} & V_{76} & \tilde{\epsilon}_7 & \frac{\sqrt{3}}{4}(V_{dd\sigma} -V_{dd\delta}) \cr 
 -\frac{1}{2}V_{pd\sigma} & 0 & 0 & 0 & 0 & 0 & \frac{\sqrt{3}}{4}(V_{dd\sigma} -V_{dd\delta}) & \epsilon_8 
\end{pmatrix}\;.
\end{align}
Here, $V_{pd\sigma}$, $V_{pd\pi}$, $V_{pd\delta}$ are Slater-Koster parameters obtained from Wannier90 downfolding corresponding to the couplings between Cr-$d$ and Te-$p$ states. Similarly, $V_{dd\sigma}$ and $V_{dd\delta}$ denote the direct coupling between the left and right Cr-$d$ states. They account for the antiferromagnetic couplings. 

$\phi_9$ and $\phi_{10}$ couple with $\phi_5$, $\phi_6$, and $\phi_7$, further renormalizing their energies and inducing effective couplings. $\tilde{\epsilon}_5$, $\tilde{\epsilon}_6$, $\tilde{\epsilon}_7$, and the emerging coupling $V_{57}, V_{67}$, $V_{75}, V_{76}$ can be calculated from the following coupling matrix
\begin{align}
\begin{pmatrix}
\epsilon_5 & 0 & 0 & 0 & 0 & V_{pd\pi} \cr
0 & \epsilon_6 & 0 & V_{pd\pi} & V_{pd\pi} & 0 \cr
0 & 0 & \epsilon_7 & \frac{\sqrt{3}}{2}V_{pd\sigma} & 0 & -\frac{1}{2}V_{pd\sigma} \cr
0 & V_{pd\pi} & \frac{\sqrt{3}}{2}V_{pd\sigma} & \epsilon_9 & 0 & 0 \cr
0 & V_{pd\pi} & 0 & 0 & \epsilon_{10} & 0 \cr
V_{pd\pi} & 0 & -\frac{1}{2}V_{pd\sigma} & 0 & 0 & \epsilon_{11} 
\end{pmatrix}\;.
\end{align}
After removing the last two rows and columns, one can easily obtain the effective Hamiltonian matrix for states $\phi_5, \phi_6$, and $\phi_7$.  
\begin{align}
\begin{pmatrix}
\tilde{\epsilon}_5 & 0 & V_{57} \cr
0 & \tilde{\epsilon}_6 & V_{57} \cr
V_{75} & V_{76} & \tilde{\epsilon}_7 
\end{pmatrix}
=\begin{pmatrix}
\epsilon_5 & 0 & \cr
0 & \epsilon_6 & 0 \cr
0 & 0 & \epsilon_7
\end{pmatrix}
+
\begin{pmatrix}
0 & 0 & V_{pd\pi} \cr
V_{pd\pi} & V_{pd\pi} & 0 \cr
\frac{\sqrt{3}}{2}V_{pd\sigma} & 0 & -\frac{1}{2}V_{pd\sigma}
\end{pmatrix}
\cdot
\begin{pmatrix}
E - \epsilon_9 & 0 & 0 \cr
0 & E - \epsilon_{10} & 0 \cr
0 & 0 & E-\epsilon_{11}
\end{pmatrix}^{-1}
\cdot
\begin{pmatrix}
0 & V_{pd\pi} & \frac{\sqrt{3}}{2}V_{pd\sigma} \cr
0 & V_{pd\pi} & 0 \cr
V_{pd\pi} & 0 & -\frac{1}{2}V_{pd\sigma}
\end{pmatrix}
\end{align}

To decouple the excited states $\phi_{2\cdots8}$ from $\phi_1$, we eliminate the matrix elements $H_{1i}$ for $i=2,\cdots, 8$ by using the $2,\cdots, 8$ rows, which is equivalent to the second-order perturbation after truncating $E_{\uparrow\uparrow}$ to the first order.   
\begin{align}
E_{\uparrow\uparrow} & \approx \epsilon_1 + \frac{3}{4}\frac{V_{pd\sigma}^2}{E - \epsilon_2 } + \frac{3}{16}\frac{(V_{dd\sigma}-V_{dd\delta})^2}{E - \epsilon_3} + (H_{14}, \cdots , H_{18}) \nonumber \\ 
&\cdot\begin{pmatrix}
E - \epsilon_4 & -V_{pd\pi} & 0 & 0 & 0 \cr
-V_{pd\pi} & E - \epsilon_5 - \frac{V_{pd\pi}^2}{E - \epsilon_{11}} & 0 & -\frac{1}{2}\frac{V_{pd\pi}V_{pd\sigma}}{E- \epsilon_{11}} & 0 \cr
0 & 0 & E - \epsilon_6 - V_{pd\pi}^2\left(\frac{1}{E - \epsilon_9}+\frac{1}{E-\epsilon_{10}}\right) & \frac{\sqrt{3}}{2}\frac{V_{pd\pi}V_{pd\sigma}}{E - \epsilon_9} & 0 \cr
0 & -\frac{1}{2}\frac{V_{pd\pi}V_{pd\sigma}}{E - \epsilon_{11}}  & \frac{\sqrt{3}}{2} \frac{V_{pd\pi}V_{pd\sigma}}{E - \epsilon_{9}} &  E - \epsilon_{7} - \frac{V_{pd\sigma}^2}{4}(\frac{1}{E - \epsilon_9} + \frac{1}{E - \epsilon_{11}}) & -\frac{\sqrt{3}}{4} (V_{dd\sigma} - V_{dd\delta}) \cr
0 & 0 & 0 &   -\frac{\sqrt{3}}{4} (V_{dd\sigma} - V_{dd\delta})  & E - \epsilon_8
\end{pmatrix}^{-1}\cdot
\begin{pmatrix}
H_{41} \cr
\vdots\cr
\vdots \cr
\vdots \cr
H_{81} 
\end{pmatrix}\nonumber \\
&\approx\epsilon_1
+\frac{3}{4}\frac{V_{pd\sigma}^2}{\epsilon_1-\epsilon_2}
+\frac{1}{4}\frac{V_{pd\sigma}^2}{\epsilon_1-\epsilon_5}
+\frac{1}{4}\frac{V_{pd\sigma}^2V_{pd\pi}^2}{\left(\epsilon_1-\epsilon_5\right)^2}\left(\frac{1}{\epsilon_1-\epsilon_{11}}+\frac{1}{\epsilon_1-\epsilon_4}\right)
+\frac{3}{4}\frac{V_{pd\sigma}^2}{\epsilon_1-\epsilon_6}
+\frac{1}{4}\frac{V_{pd\sigma}^2}{\epsilon_1-\epsilon_8}
+\frac{3}{4}\frac{V_{pd\sigma}^2V_{pd\pi}^2}{\left(\epsilon_1-\epsilon_6\right)^2}\left(\frac{1}{\epsilon_1-\epsilon_9}+\frac{1}{\epsilon_1-\epsilon_{10}}\right)
\nonumber\\ 
&
+\frac{3}{16}\frac{\left(V_{dd\sigma}-V_{dd\delta}\right)^2}{\epsilon_1-\epsilon_3}
+\frac{3}{16}\frac{\left(V_{dd\sigma}-V_{dd\delta}\right)^2}{\epsilon_1-\epsilon_4}
+\frac{V_{pd\pi}^2}{\epsilon_1-\epsilon_7}
+\frac{1}{4}\frac{V_{pd\sigma}^2V_{pd\pi}^2}{\left(\epsilon_1-\epsilon_7\right)^2}\left(\frac{1}{\epsilon_1-\epsilon_9}+\frac{1}{\epsilon_1-\epsilon_{11}}\right)
-\frac{\sqrt3}{4}\frac{V_{pd\sigma}V_{pd\pi}\left(V_{dd\sigma}-V_{dd\delta}\right)}{\left(\epsilon_1-\epsilon_5\right)\left(\epsilon_1-\epsilon_4\right)}
\nonumber\\
&
+\frac{1}{2}\frac{V_{pd\sigma}^2V_{pd\pi}^2}{\left(\epsilon_1-\epsilon_5\right)\left(\epsilon_1-\epsilon_7\right)\left(\epsilon_1-\epsilon_{11}\right)}
+\frac{3}{2}\frac{V_{pd\sigma}^2V_{pd\pi}^2}{\left(\epsilon_1-\epsilon_6\right)\left(\epsilon_1-\epsilon_7\right)\left(\epsilon_1-\epsilon_9\right)}
-\frac{\sqrt3}{4}\frac{V_{pd\sigma}V_{pd\pi}\left(V_{dd\sigma}-V_{dd\delta}\right)}{\left(\epsilon_1-\epsilon_7\right)\left(\epsilon_1-\epsilon_8\right)}
+\mathcal{O}\left(V^5\right)\;.
\end{align}

From the above explicit derivation, we further summarize three GKA rules which will then be applied to other super-exchange processes to simply the calculation. 
According to the order of single-particle hopping required by the super-exchange, we categorize the above second-order corrections to the ground state energy as
\begin{itemize}
\item The two steps exchange in order of $V^2$: $V_{pd\sigma}^2$, $V_{pd\pi}^2$, $(V_{dd\sigma} - V_{dd\delta})^2$. For such processes, the energy correction can be simply written as  $\Delta E_2 = \sum_{i<j}\frac{2V_{1i}V_{i1}}{\epsilon_1-\epsilon_i}$, in which the $V_{1i}$ and $V_{i1}$ are the overlap between the ground state $|\phi_1\rangle$ and the excited state $|\phi_i\rangle$. This consists our first GKA rule. 
\item The three-step exchange in order of $V^3$: $V_{pd\sigma}V_{pd\pi}(V_{dd\sigma}-V_{dd\delta})$. The energy corrections to the ground state from such processes can be quickly written as $\Delta E_3 = \sum_{i<j}\frac{2V_{1i}V_{ij}V_{j1}}{(\epsilon_1-\epsilon_i)(\epsilon_1-\epsilon_j)}$, where the coefficient number 2 is due to the equivalence of $1\rightarrow i \rightarrow j \rightarrow 1$ and  $1\rightarrow j \rightarrow i \rightarrow 1$ exchange processes. This is our second GKA rule. 
\item The four-step exchange in order of $V^4$: $V_{pd\sigma}^2V_{pd\pi}^2$, which leads to the last GKA rule $\Delta E_4=\sum_{ijk}\frac{(2-\delta_{ik})V_{1i}V_{ij}V_{jk}V_{k1}}{(\epsilon_1-\epsilon_i)(\epsilon_1-\epsilon_j)(\epsilon_1-\epsilon_k)}$. 
\end{itemize}

With the three GKA rules, one can simplify the tedious matrix reduction to the following procedures that can be programed to automatically analyze all possible super-exchange paths. 
\begin{itemize}
\item Starting from the ground state, one first finds out all possible excited states that relate to the ground state by finite times of single-particle hopping. In this work, we use the depth-first search (DFS) algorithm~\cite{doi:10.1137/0201010} to search all possible excited states. 
\item Connect the ground state with these excited states with the single-particle hopping and restrict the initial and final states to be the same ground state, which will generate a super-exchange path. 
As the longer path gives rise to smaller energy corrections, we restrict the exchange path to be no longer than 4 steps of exchange. 
In this way, we find all allowed super-exchange paths and calculate the state-overlap appeared in the path. The state-overlap is parametrized with the Slater-Koster integrals. 
We note that, in addition to the Slater-Koster parametrization, one also needs to be careful with the overall sign-change caused by the exchange of electron pairs. Whenever an electron switches position with another electron, there appears a minus sign. Accumulating all minus signs generated in the super-exchange path, one gets the overall sign-change of this path. 
\item Following the three GKA rules summarized above, we evaluate the energy corrections from $\epsilon_i$ and $V_i$ to the ground state for a given super-exchange path.      
\end{itemize}

\begin{figure}[htbp]
\centering
\includegraphics[width=\linewidth]{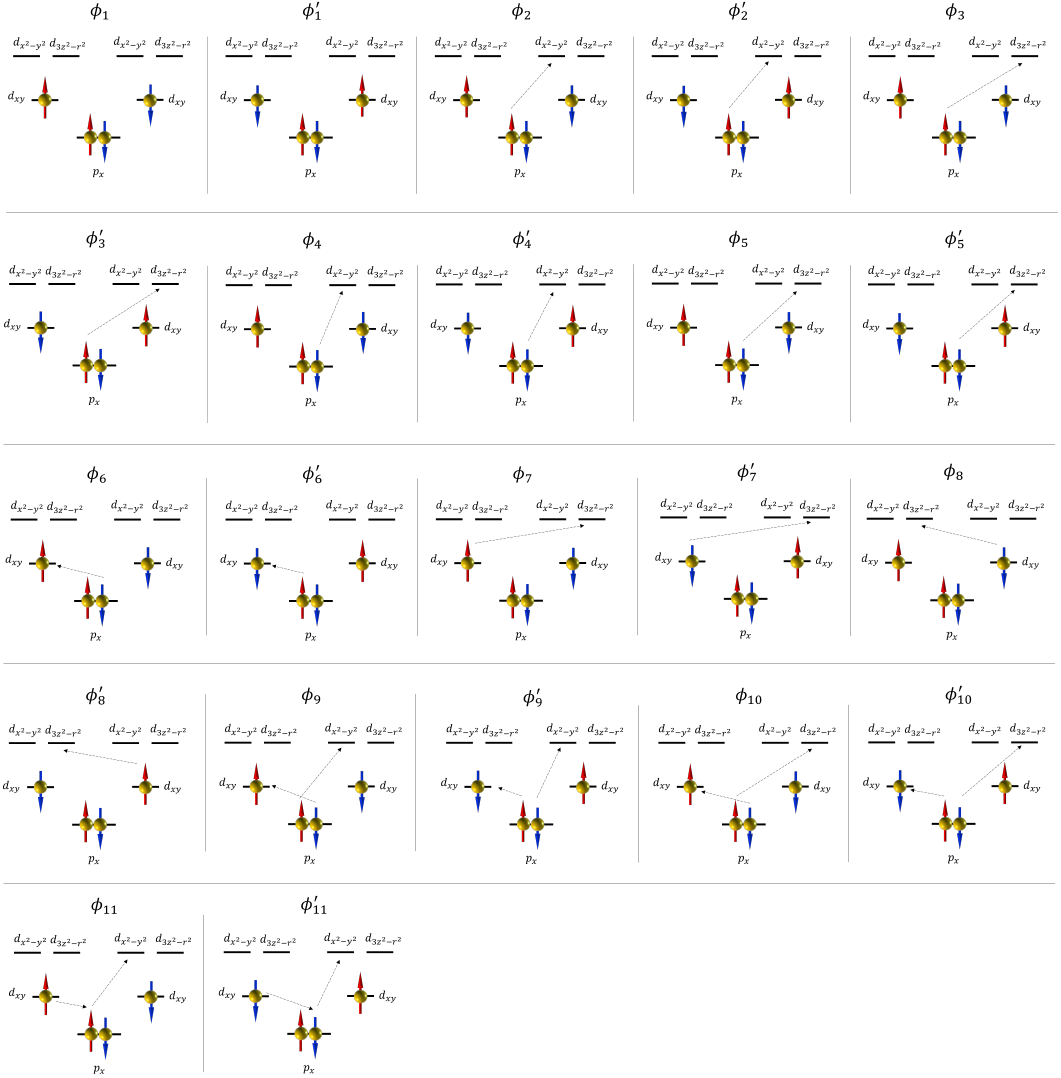}
\caption{Exchange processes for antiparallel-spin configuration.}
\label{FigS:antiparallel_1}
\end{figure}

After getting the energy correction to both the spin-paralllel and spin-antiparallel configurations of two neighboring Cr ions, we finally obtain the exchange coupling $J$ as
\begin{align}
    J(t_{2g}-p-e_g) = 2(2E_{\uparrow\uparrow}-2E_{\uparrow\downarrow})\;.
\end{align}
by assuming the isotropic Heisenberg model as the effective model for the spin excitations in insulating CrGeTe$_{3}$ (without considering the spin-orbital coupling). 
The additional factor 2 in the brackets is due to the fact that, in addition to the participation of the $p_x$ orbital, there is also a symmetrical exchange involving the $p_y$ orbital alone in this plane, and its energy is the same as that from the $p_x$ orbital. 

\begin{figure}[htbp]
\centering
\includegraphics[width=0.6\linewidth]{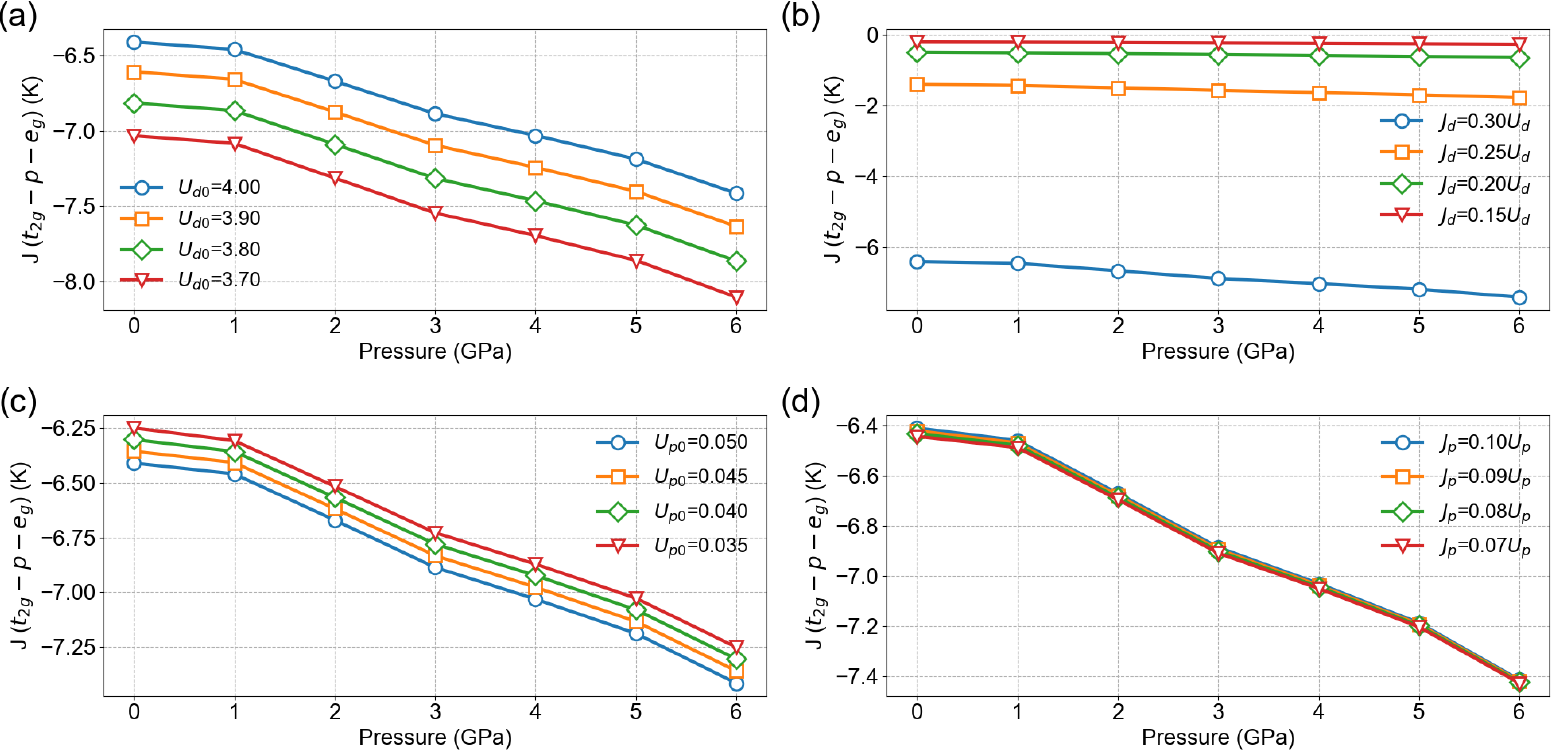}
\caption{Effective exchange coupling contributed by super-exchange path $t_{2g}-p-e_g$}
\label{FigS:tpe}
\end{figure}

The Slater-Koster parameters are obtained from Wannier projection of the DFT electronic structure with all Cr-$d$ and Te-$p$ orbitals. 
We extract the onsite energies and hopping amplitudes from the local Hamiltonian matrix of the projected tight-binding model. 
To systematically account for the evolution of Slater-Koster parameters and the electronic correlations with pressure, we set $V_{dd\sigma}=10V_{dd\pi}=10V_{dd\delta}=0.01$, $V_{pd\sigma}=100V_{pd\pi}=1.0+0.03P$, $U_d=U_{d0}-0.05P$, $J_d=0.3U_d$, $U_p=U_{p0}-0.005P$, $J_p=0.1U_p$ for pressure $P=0,\ 1,\ \cdots,\ 6$ GPa. 
$U_{d0}$ and $U_{p0}$ are interaction parameters used at ambient pressure. 
The effective exchange coupling obtained for $t_{2g}-p-e_g$ is shown in Fig.~\ref{FigS:tpe}, where $J$ is negative indicating the ferromagnetic contribution of this super-exchange process.
The four curves in each plot correspond to different interaction parameters. 
In Fig.~\ref{FigS:tpe}(a), with the increase of $U_{d0}$, $J$ becomes smaller owing to the fact that $U_{d}$ always appears in the denominator of the energy corrections in the $t_{2g}-p-e_{g}$ process.  
This path provides the main contribution to ferromagnetism. 
While, as the Hund's coupling $J_{d}$ increases, the ferromagnetism contributed by this path becomes stronger. 
This is because the Hund's energy prefers the highest spin state and aligns electrons parallelly in different orbitals. 
Fig.~\ref{FigS:tpe}(c) and (d) show that $J$ is nearly unchanged by $U_p$ and $J_p$. As there is only one Te-$p$ orbital participating in this exchange path,  the Hund's energy which act on electrons at different orbitals, thus, has no effect in this case.

\subsection{Super-exchange process $e_{g}-p_{1}|p_{2}-e_{g}$}
\begin{figure}[htbp]
\centering
\includegraphics[width=0.6\linewidth]{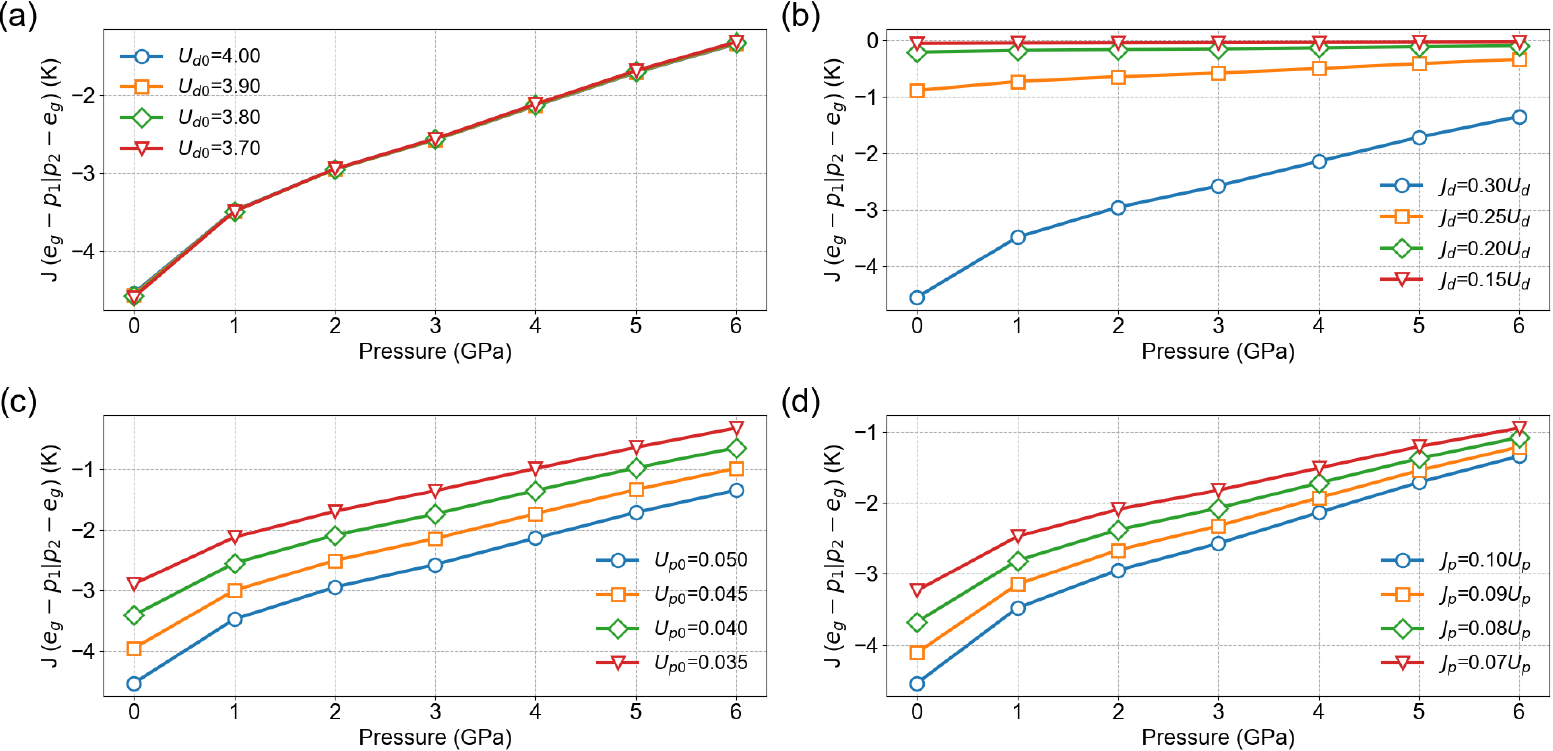}
\caption{Effective exchange coupling contributed by super-exchange path $e_{g}-p1|p2-e_g$}
\label{FigS:eppe}
\end{figure}

Similarly, we also evaluated the exchange coupling for super-exchange process $e_{g}-p_{1}|p_{2}-e_{g}$ which involves two different Te-$p$ orbitals.
The effective exchange coupling evolution with pressure for this super-exchange process is displayed in Fig.~\ref{FigS:eppe}, which contributes ferromagnetically to the overall $J$. 
The most significant observation is that the $J$ value contributed by this path is almost unaffected by $U_d$. 
As $e_g$ orbital is empty and this path does not involve $t_{2g}$ orbital, new doubly occupied states are rarely generated by this super-exchange within the limited number of exchange steps, and, once generated, they correspond to high energy states with little contribution to the overall energy correction. 
As a consequence, $U_d$ rarely contributes and electrons often occupy different orbitals in the same Cr ion. 
The local Hund's coupling prefers a parallel alignment of their spins, leading to the preference on the ferromagnetic coupling. 
For the same reason, electrons travel from the other Cr ion will also like to align parallelly with the spins at this Cr ion.
Thus, with the increase of $J_d$, ferromagnetic coupling is also enhanced as shown in Fig.~\ref{FigS:eppe}(b). 

Similarly, due to the Hund's coupling of the $p$ orbitals, for the parallel spin alignment of the neighboring $t_{2g}$ electrons,  only the opposite spins from the $p$ orbitals can travel to them leaving the two electrons remaining at the Te site parallel in spins. 
Otherwise, for the antiparallel configuration of the neighboring $t_{2g}$ electrons, after hopping two $p$ electrons to Cr ions, the remaining two $p$ electrons will be antiparallelly aligned, which is higher in energy than the ferromagnetic configuration due to the Hund's coupling of $J_p$. 
Consequently, $J_p$ also favors the ferromagnetic alignment of Cr $d$ electrons. Increasing $U_p$ and $J_p$ results in the enhancement of ferromagnetic coupling, as shown in Fig.~\ref{FigS:eppe}(c) and (d).

\subsection{Super-exchange process $t_{2g}-p_{1}|p_{2}-t_{2g}$}
\begin{figure}[htbp]
\centering
\includegraphics[width=0.6\linewidth]{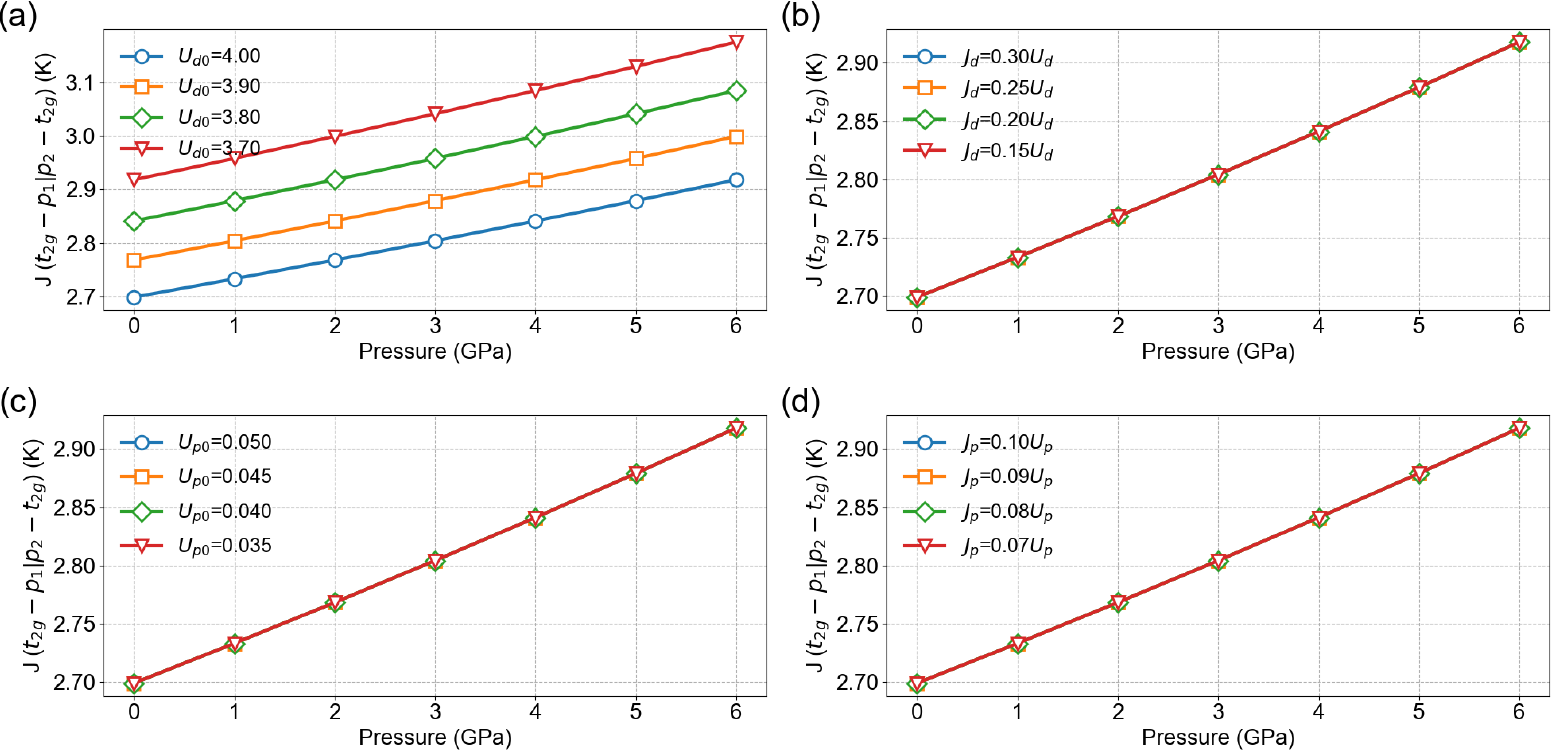}
\caption{Effective exchange coupling contributed by super-exchange path $t_{2g}-p_{1}|p_{2}-t_{2g}$}
\label{FigS:tppt}
\end{figure}

$t_{2g}-p_{1}|p_{2}-t_{2g}$ corresponds to the super-exchange process involving only $t_{2g}$ orbitals and the ligand $p$ orbitals. No $e_{g}$ state participates in this process. 
The exchange coupling contributed by this process is shown in Fig.~\ref{FigS:tppt}. %
As shown in this figure, $t_{2g}-p_{1}|p_{2}-t_{2g}$ contributes antiferromagnetically to the exchange coupling $J$. 
This process contains the contribution from a direct $t_{2g}$ electron hopping between two neighboring Cr ions, which leads to an antiferromagnetic coupling between them. 
Although the overall amplitude of such hopping is small, it increases with increasing pressure. 
Additionally, the local Coulomb repulsion $U_d$ decreases with pressure, further reducing the energy barrier for two Cr-$t_{2g}$ electrons to occupy the same state. 
However, since only the $t_{2g}$ orbital of Cr is involved in this supe-rexchange path, there is no interaction between electrons in different $d$ orbitals at the same site, so the size of the Hund's energy $J_d$ has no effect on this path, as shown in Fig.~\ref{FigS:tppt}(b). 

Figure~\ref{FigS:tppt}(c) and (d) also indicate that the exchange coupling from this process is independent of $U_{p}$ and $J_p$.  
For the contribution from the direct exchange of Cr $d$ electrons, the $p$ state occupancy remains unchanged during exchange. 
Consequently, the energy correction from such process has nothing to do with $U_p$ and $J_p$.
For processes with only one $p$ electron participating in the exchange, the remaining occupancy of $p$ states is always one singly occupied and one doubly occupied state irrespective of the Cr $d$ spin configuration. 
Thus, the energy difference between the parallel and antiparallel configuration of Cr $d$ states does not change with the change of $U_p$ and $J_p$. 
As for the exchange processes involving two $p$ electrons, which concerns at least four state overlaps, the amplitude of such processes is too small to observe the influence of $U_p$ and $J_p$. 
Thus, the direct exchange dominated antiferromagnetic coupling is nearly independent of $J_d$, $U_p$, and $J_p$ from our calculations. 

Collecting all the exchange processes discussed above, we obtained an overall ferromagnetic coupling $J$. 
With the increase of pressure, the ferromagnetic exchange coupling contributed by the $t_{2g}-p-e_g$ super-exchange path and the antiferromagnetic $t_{2g}-p_1|p_2-t_{2g}$ path increase, while the ferromagnetic $e_g-p_1|p_2-e_g$ path gradually decreases. 
Consequently, the effective ferromagnetism coupling decreases with increasing pressure in the low-pressure region, which is consistent with the reduction of Curie temperature observed in experiment.

\section{Evidence of persisting local moments in metallic phase}
\begin{figure}[htbp]
\centering
\includegraphics[width=0.6\linewidth]{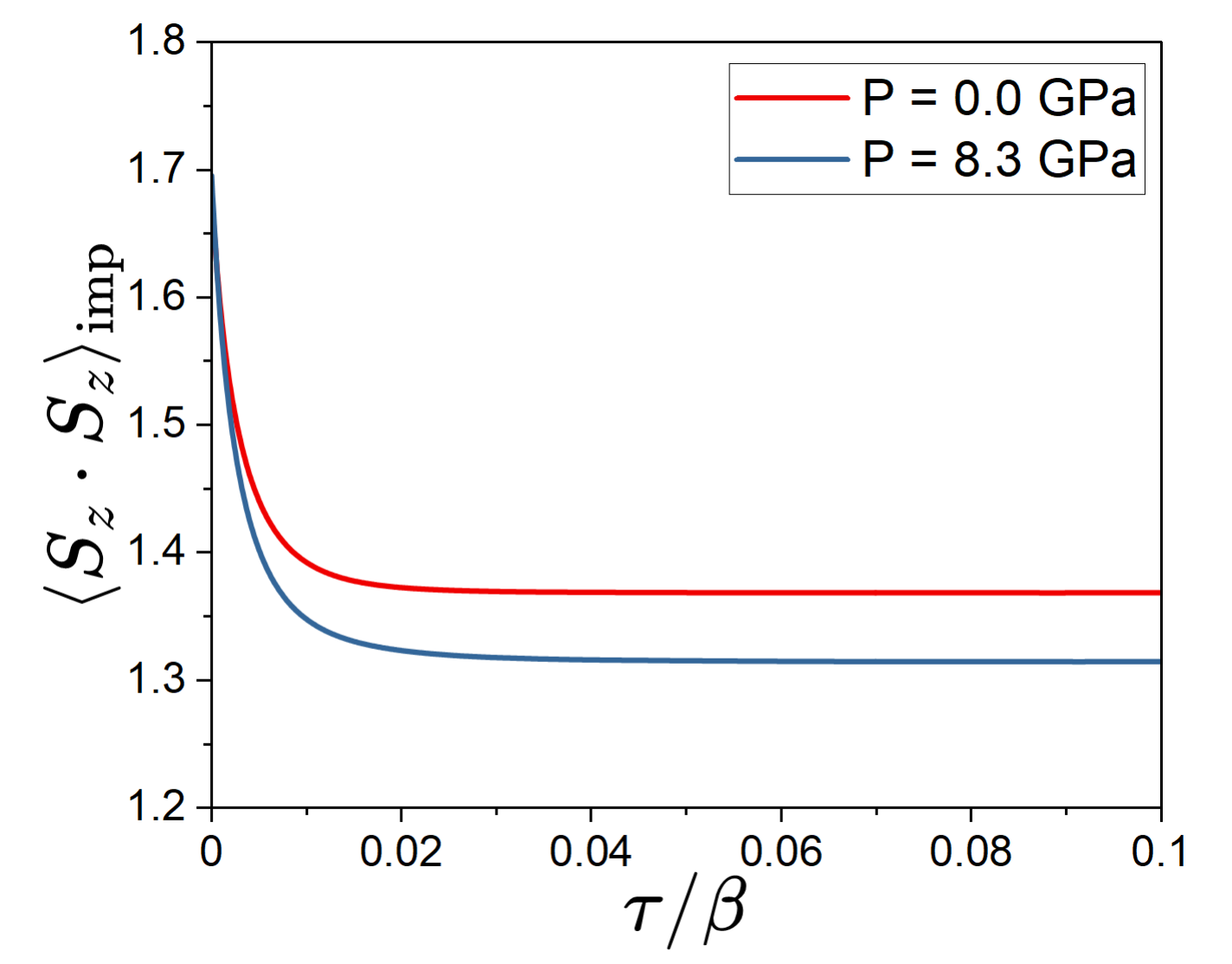}
\caption{Local spin correlation function as a function of $\tau$.}
\label{FigS:local_moment}
\end{figure}
To reveal local moments in the metallic phase, we further performed DFT + DMFT calculations on the local spin susceptibility, as shown in Fig.~\ref{FigS:local_moment}. The presence of local moments can be seen from the long-time behavior of $\langle S_z(\tau)S_z\rangle_{imp}$. The finite constant value of the $\langle S_z(\tau)S_z\rangle_{imp}$ at  $\tau\rightarrow\beta/2$ signals the presence of local moments. In the metallic state at P = 8.3 GPa, the saturation value of $\langle S_z(\tau)S_z\rangle_{imp}$ becomes smaller but remains finite, indicating a stronger screening from the mobile electrons in the metallic state. The persisting local moments, the enhanced metallicity, and the emerging kinetic-exchange in the metallic phase give rise to a joint contribution to the enhanced ferromagnetism in CrGeTe3. Specifically, we believe that the leading contribution is from ferromagnetic kinetic-exchange. Additionally, the Stoner instability of the enhanced DOS and the possible remaining super-exchange mechanism further strengthen the metallic ferroamgnetism.

\section{Estimation of interaction parameters under pressure}

we also performed constrained random phase approximation (cRPA) calculations within VASP for pressures at 0 – 4 GPa. The plane-wave cut-off energy was set to 400 eV, and the k-mesh was set to 6×6×6. The total number of bands considered in the self-consistent calculations is 192 and wannierization was done for Cr-$d$, Ge-$p$, O-$p$ states of totally 34 orbitals, which was then used as screening to the lowest 10 Cr-$d$ bands. Thus, the low-energy screening is from Ge-$p$ and O-$p$ in this calculation, and the high-energy screening is from the rest of 192-34=158 Bloch states. The interaction parameters for the Cr-d electrons are summarized below in Tab. I. 
 
 \begin{figure}[htbp]
 \centering
 \includegraphics[width=0.8\linewidth]{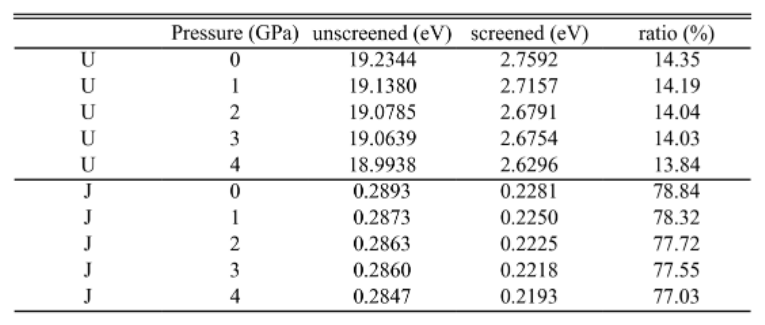}\\
 Table I: Coulomb interaction $U$ and Hund’s coupling $J$  estimated from constrained random-phase approximation (cRPA) calculation.
 \end{figure}
 
The bare Coulomb interactions on Cr-d electrons is about 19 eV, which is significantly reduced to about 2.7 eV after screening. But the Hund’s coupling is much less affected by screening as usual. As evident from the table, as the increase of pressure, both Coulomb and Hund’s parameters become smaller as we discussed in the manuscript. Together with the detailed change of crystal structure, the reduced effective interaction explains the insulator-metal transition without structure transition of CrGeTe3.

\end{document}